Technical Whitepaper

September 2024

# XDC Gasless Subnet: Gasless Subnet Staking dApp *for*

## XDC NETWORK

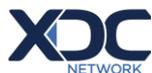

Mohuya Chakraborty
Atul Khekade

# XDC Gasless Subnet: Gasless Subnet Staking dApp for XDC Network

Mohuya Chakraborty (mohuyacb@gmail.com)

Atul Khekade (atul@xinfin.org)

**Abstract.** With a delegated proof-of-stake (XDPoS) consensus mechanism, the XDC Network is an enterprise-focused blockchain platform that combines the strength of public and private blockchains to provide quick transaction times, low energy consumption, and economical gas fees. XDC is designed for interoperability and supports decentralized apps (dApps) and integrates smoothly with financial systems. It is perfect for trade financing and tokenisation of physical assets because of its emphasis on security and scalability. However, there are a few critical issues that hamper wider acceptance and usability for certain high-frequency applications. This whitepaper introduces a novel and enthralling dApp for establishing a gasless subnet in which mainnet XDC can be staked to spin off a subnet that functions similarly to a non-crypto network, accepting currency fees on the XDC network. This would allow users to stake their tokens without incurring gas fees making the staking process more efficient, cost-effective, and simultaneously enhancing scalability. Performance evaluation of the dApp shows promising results in terms of throughput, latency, scalability, security, and cost efficiency. The use cases and applications of this approach along with challenges and ensuing solutions are included.

## Table of Contents

# 1 Introduction

Blockchain technology has transformed various industries by enabling secure, transparent, and decentralized transactions. Among the leading blockchain ecosystems, the XDC network stands out for its focus on enterprise-ready hybrid blockchain solutions. With its energy-efficient consensus mechanism and low transaction fees, XDC is optimized for applications in trade finance, supply chain management, decentralized applications (dApps), and decentralized finance (DeFi). However, even with these advantages, several critical issues hinder broader adoption and usability for certain high-frequency applications.

One of the key challenges is transaction cost scalability. While the XDC Network already offers low fees compared to other blockchain platforms, applications that require frequent interactions—such as microtransactions, IoT-based networks, and gaming—face a cumulative burden from these costs. The problem becomes particularly acute when end users, who may be unfamiliar with cryptocurrency mechanisms, are required to manage and pay for gas fees. This creates friction, reducing usability and limiting adoption. Additionally, incentivizing validators in a sustainable way without inflating the token supply poses another challenge. Traditional staking models often rely on inflationary rewards to compensate validators, which can undermine the long-term economic stability of the network. As more subnetworks (or subchains) are spun off, balancing security, decentralization, and validator incentives becomes more difficult without relying on inflation. Scalability is another pressing concern. As the XDC network grows and more dApps are deployed, maintaining the same low transaction costs and high throughput across multiple subnets becomes increasingly difficult. The need for a model that supports secure, high-volume transactions without overloading the network or inflating transaction fees is crucial.

One solution gaining attention is the development of gasless subnets—dedicated subnetworks that enable users to perform transactions without incurring the traditional gas fees associated with blockchain operations. For the XDC network, a gasless architecture would significantly enhance its usability, especially for dApps that require frictionless user experiences, such as IoT-based networks, gaming, and microtransactions. However, ensuring network security, validator participation, and decentralization in gasless subnets presents new challenges, particularly in systems that avoid inflationary token models for incentivization.

In response to these challenges, this paper proposes a novel gasless subnet-level staking mechanism specifically designed for the XDC network. This mechanism allows subnet operators to stake XDC mainnet tokens to establish and run gasless subnets. It directly addresses the problem of transaction costs by eliminating the need for users to pay gas fees, improving the overall user experience and making the network more accessible to non-crypto-savvy participants. Furthermore, validators are compensated not through inflation but via transaction fee redistribution or other incentive structures, ensuring long-term sustainability and security. This approach solves the problem of high transaction costs, offers a non-inflationary reward model, and supports scalability by creating gasless subnets without compromising on decentralization or network security.

It aligns with the XDC network's goals of providing scalable, enterprise-grade solutions and has the potential to revolutionize how applications interact with the network. It offers a way for subnet operators to support high-throughput applications without burdening users with transaction fees, further driving adoption in key sectors.

The main contributions of the paper include:
1. Creation of subnet architecture
2. Implementation of gasless transactions
3. Fiat fee integration
4. Performance analysis of the proposed model

This is how the remainder of the paper is structured. [Section 2](#) reviews related works in solving scalability issues of blockchain networks in general. [Section 3](#) gives an overview of the proposed solution including the architecture, design principles, implementation details, and performance evaluation. [Section 4](#) is devoted to the use cases and applications, whereas [Section 5](#) highlights the challenges of the proposed solution and their remedies. The roadmap and the team behind this work are presented in [Section 6](#) and [Section 7](#) respectively, Finally, [Section 8](#) concludes with a discussion on the potential impact of this innovation on the XDC ecosystem and its broader applications.

## 2    Related Works

Several works have reported widespread use of sharding, subnets, off-chain scaling, layer-2 scaling, consensus mechanisms, and alternative approaches in several blockchain networks for solving scalability issues that in turn further lead to transaction delays, excessive energy consumption, low throughput thereby compromising efficiency [1]. Reduction in gas cost through proper design of blockchain smart contract for educational certificate generation and storage has been reported in [2]. In [3], the authors have discussed an architecture to develop a trustless novel gasless on-chain password manager including a robust recovery mechanism with a fundamental goal to overcome the prevalent challenges related to gas fees in current on-chain password managers, which often impede accessibility and usability. Subnet scaling solution aims to address scalability challenges in blockchain networks.

Subnets are decentralized networks within a blockchain ecosystem that operate independently and may have their own set of rules, validators, and consensus mechanisms. They are designed to enhance scalability and improve performance by enabling parallel processing of transactions and smart contracts [4]. There are several subnets in the industry at present including Avalance subnets, Cosmos SDK Subnets, Polkadot Parachains, Near Nightshade Shards, Fantom Opera Chains etc. Gasless meta transactions have been in vogue for a long time now in the blockchain ecosystem. Normally, when a user wants to do a transaction on blockchain, he has to pay a fee called gas. Gas fees are calculated based on the gas limit (the maximum amount of gas an user is willing to spend) and the gas price (the amount of token (viz., XDC) an user is willing to pay per unit of gas). The formula is given by:

$$\text{Gas Fee} = \text{Gas Limit} \times (\text{Base Fee} + \text{Priority Fee}) \text{--------(1)}$$

But with gasless meta transactions, users do not need to pay this fee, which makes it easier and cheaper for them to use the blockchain and participate in transactions [5]. Integrating the concept of subnet (Layer-2) and gasless meta transactions, this paper proposes a solution to further enhance the throughput, latency, scalability, and efficiency of the XDC network.

## 3  Proposed Solution

This section discusses the overall architecture, design principles, implementation, and performance evaluation of the proposed approach.

### 3.1  Architecture

To create a gasless XDC subnet where mainnet XDC can be staked to spin off a subnet that operates like a non-crypto network with fiat fees, several components are leveraged based on the XDC network's capabilities and external solutions are integrated for handling fiat transactions. The architecture of the proposed gasless subnet-level staking mechanism in the XDC network involves several key components to ensure smooth operation without requiring transaction fees (gas) for subnet participants as depicted in Figure 1.

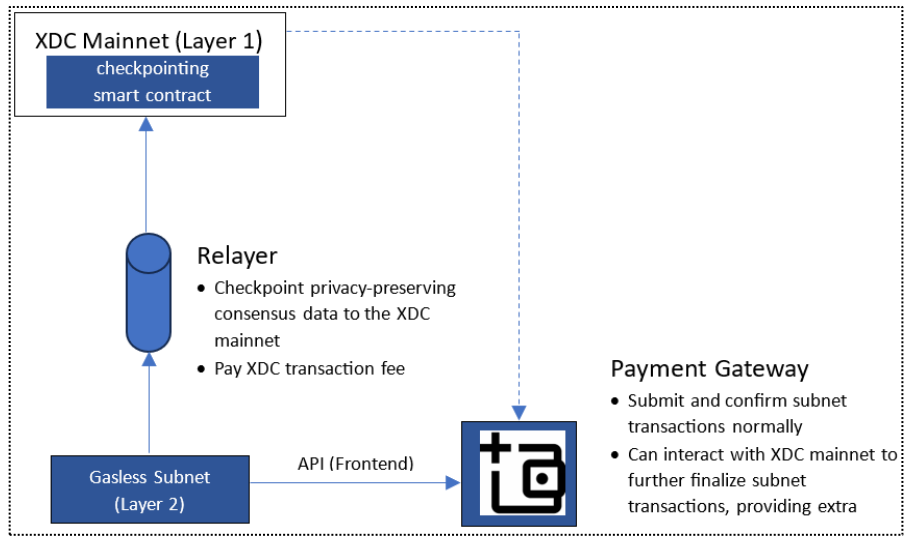

Figure 1 Gasless Subnet Level Staking concept

There are four major components: XDC Mainnet (Layer 1), Gassless Subnet (layer 2), Relayer, and Payment Gateway. The XDC mainnet is the primary blockchain where XDC tokens reside and core network functions (such as validation, staking, and transactions) occur. It serves as the backbone of the entire network, ensuring security and consensus. The mainnet maintains interaction with subnets for transaction finality, validation, and data security. Checkpoints or state updates from the subnet are periodically posted to the mainnet for integrity and cross-chain validation. The mainnet XDC can be staked to spin off a subnet that operates like a non-crypto network with fiat fees

The gasless subnet is a secondary layer (Layer 2) built on top of the mainnet that processes transactions without requiring gas fees. It is optimized for specific use cases (e.g., gaming, DeFi) where users can transact without needing to hold cryptocurrency for fees. The key feature is that users do not need to pay gas fees for each transaction. Instead, an alternative fee system (e.g., fiat payments or other rewards mechanisms) is used to cover operational costs.

Masternodes validate transactions and ensure consensus within the subnet. They do not charge gas for transactions, relying instead on other incentives like rewards from staking, fee redistribution, or fiat-based payments. Standby nodes wait to take over in case of failures or when more capacity is required. Even when idle, these nodes may earn rewards, ensuring all participants benefit. In a gasless system, a relayer is an intermediary that facilitates the submission of user transactions. Since users do not pay gas, the relayer takes responsibility for forwarding transactions to the mainnet or validating them within the subnet. Relayers can batch transactions or handle them on behalf of users and validators, reducing the complexity for end-users. Relayers may be compensated through alternative methods such as fees paid in fiat or by being allocated a portion of the rewards generated from staking or transaction fees.

The payment gateway serves as the bridge between the blockchain ecosystem and traditional financial systems. In a gasless subnet, it allows users to interact with the blockchain without holding or using cryptocurrency for gas fees. Users can pay fees in fiat currencies (like USD, EUR) instead of using XDC tokens or other cryptocurrencies. The payment gateway converts these fiat fees into operational costs for the network, ensuring that validators, masternodes, and relayers are compensated. The payment gateway enables the network to attract non-crypto users by removing the barrier of needing to understand or use cryptocurrency, making the network more user-friendly and accessible.

These components work together to maintain a secure, efficient, and user-friendly blockchain experience in the XDC network's gasless subnet architecture, providing scalability while reducing entry barriers for users.

## 3.2 Design Principles

The design principles include designing the subnet architecture, writing the gasless meta-transaction smart contract, checking the security and compliance, scheming the fiat fee integration through payment gateway using a fronend dApp that should run on a server followed by testing and deployment into the XDC network. The details of the design principles are enumerated below.

### 3.2.1 Designing the Subnet Architecture

*Subnet Deployment:* Using the XDC network's ability to support multiple chains, custom subnets that operate under their own consensus rules adapted for fiat fee processing as depicted in Figure 2 may be created.

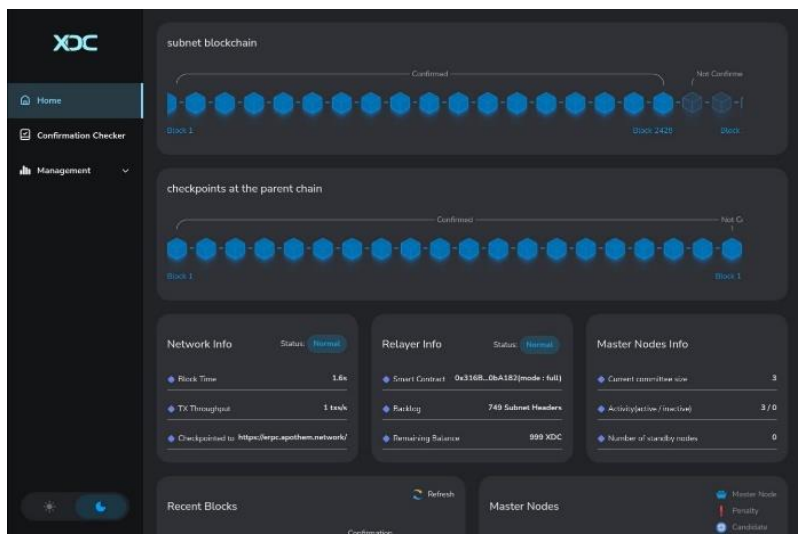

Figure 2 Subnet Creation

*Staking Mechanism:* Implementing a staking contract on the XDC mainnet where users can lock their XDC tokens to receive access or governance rights on the new subnet.

### 3.2.2 Implementing Gasless Transactions

*Meta-Transactions:* Using meta-transactions to enable gasless transactions within the subnet involves transactions being sponsored by a relayer who pays the gas fee in XDC, while the end-user incurs the cost in fiat.

*Gas Station Network (GSN):* Integrating the Gas Station Network to manage meta-transactions by modifying the GSN to accept XDC for transaction fees and reimbursing relayers in XDC, while linking charges to users in fiat.

### 3.2.3 Fiat Fee Integration

*Fiat Gateway:* Partnering with a fiat payment processor (viz., Square) to facilitate fiat payments within the subnet. This gateway would handle conversion rates between fiat and XDC for fee calculation.

*Fee Management Contract:* Developing a smart contract to manage the conversion of fiat fees to XDC within the subnet, ensuring that relayers or validators are compensated appropriately in XDC.

### 3.2.3 Security and Compliance

*Smart Contract Audits:* Conducting thorough audits of all smart contracts involved, particularly those handling staking, meta-transactions, and fiat conversions to ensure security and compliance with relevant regulations.

*KYC/AML Compliance:* Integrating KYC/AML processes for users engaging in fiat transactions within the subnet to adhere to financial regulations.

### 3.2.5 Testing and Deployment

*Apothem:* Deploying the subnet initially on a testnet (Apothem) to trial the entire system with test tokens and simulated fiat transactions.

*Mainnet Launch:* After successful testing and auditing, launching the subnet on the XDC mainnet. Ensuring that robust monitoring and support systems are in place to address any issues.

## 3.3 Implementation Details

To support gasless transactions on the XDC network, we have implemented a mechanism that leverages meta-transactions. Meta-transactions allow transactions to be submitted without the user needing to pay for the gas fees directly. Instead, a relayer pays the gas fees and can later be reimbursed in the transaction's token or via other agreed-upon methods. The details of setting up gasless transactions using meta-transactions are as follows.

- *Smart Contract Modification:* Modify or deploy smart contracts that are capable of handling meta-transactions. These contracts typically use signatures to authenticate users without requiring them to spend gas. Users sign a transaction with their private key off-chain and send this signature to a relayer.
- *Relayer Set Up:* A relayer is a service (or another user) that submits the signed transaction to the blockchain. This relayer pays the gas fees but can be compensated either through a fee included in the transaction, via separate payment channels, or through application-specific tokens.
- *Validate Signature:* The smart contract validates the signature to confirm that the transaction is indeed authorized by the signer. This involves reconstructing the signed message from the signature and comparing it against the signer's address.
- *Execute Transaction:* Once the signature is validated, the transaction is executed as if it was initiated by the original sender, even though the relayer is submitting it.
- *Compensate the Relayer:* The compensation model for the relayer can vary. Some common models include charging the user in application-specific tokens, adding a fee to the transaction that goes to the relayer, or through agreements external to the blockchain.
- *Integration with Frontend:* The frontend collects the user's signature using a web3 library (like ethers.js). The signed message is then sent to a relayer. The relayer submits the transaction to the XDC network. Using such a setup, you can effectively enable gasless transactions on the XDC network, enhancing user experience, especially for those who may not want to spend XDC tokens for gas fees or are new to blockchain interactions.

Figure 3 shows the On-Chain and Off-Chain interactions as per the protocol.

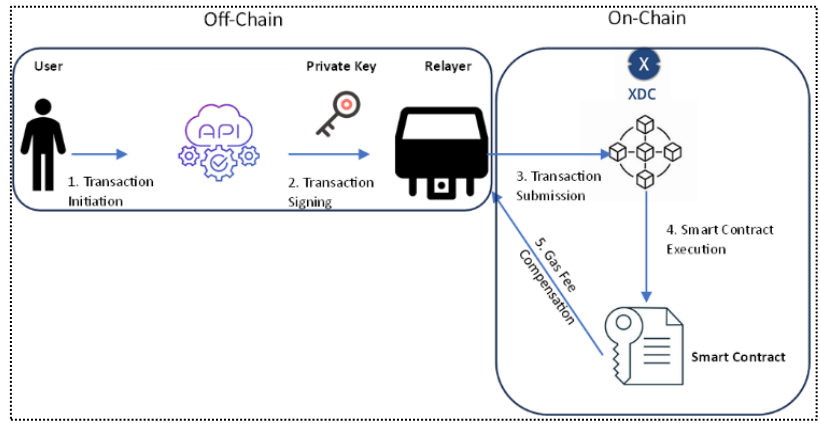

Figure 3 Off-Chain and On-Chain Interactions

The Off-Chain Interaction comprises (1) transaction initiation and (2) transaction signing.
1. *Transaction Initiation:* The user initiates a transaction off-chain, typically through a user interface or application using fiat currency.
2. *Transaction Signing:* The transaction is signed by the user off-chain using their private key and sent to a relayer, which picks up the signed transaction. The relayer is responsible for submitting the transaction to the XDC blockchain on behalf of the user.

The On-Chain Interaction comprises (3) transaction, (4) submission smart contract execution, and (5) gas fee compensation.
3. *Transaction Submission:* The relayer submits the transaction to the XDC Network. This involves paying the gas fee required for the transaction.
4. *Smart Contract Execution:* The transaction is processed by the smart contract on the blockchain. The smart contract verifies the signature and executes the transaction.
5. *Gas Fee Compensation:* The relayer can be compensated for the gas fees through various mechanisms, such as charging a small fee to the user or being reimbursed by the application provider.

Flow Diagram of integrating meta transaction into dApp frontend depicted in Figure 4, shows the entire process illustrating in detail the interactions among the actors behind the scenes. The user connects to the frontend dApp to initiate the process. It gets a payment ID from the wallet (for previously paid fees in fiat currency) and sends it to the relayer. The relayer upon getting authorization from the wallet, sends the transaction to the XDC blockchain using gas. The relayer will be rewarded later on for his task. Figure 5 depicts the Meta-Transaction dApp output deployed at localhost:3061 for two test cases (failure and success). The average execution time of the dApp has been calculated to be 18.4 sec. It includes the processing time of the payment gateway and the running time of the meta-transaction contract in XDC Apothem (Testnet) blockchain network. The project is available on Github (https://github.com/CCoE-UEMK/MetaTransactionUser).

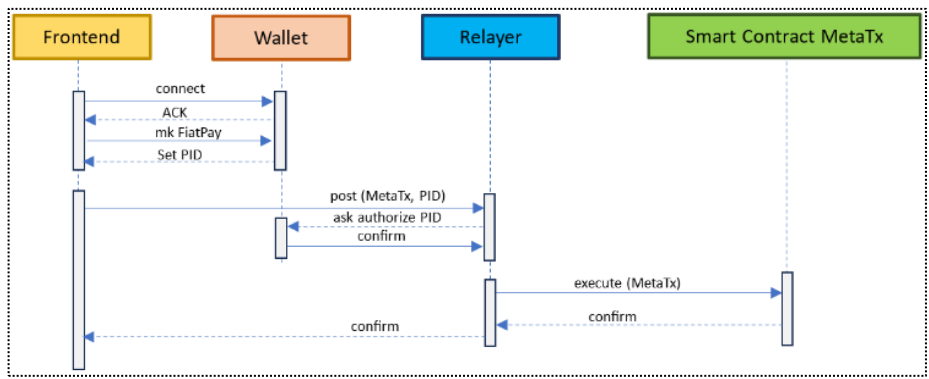

Figure 4 Integrating Meta Transaction into XDC dApp Frontend

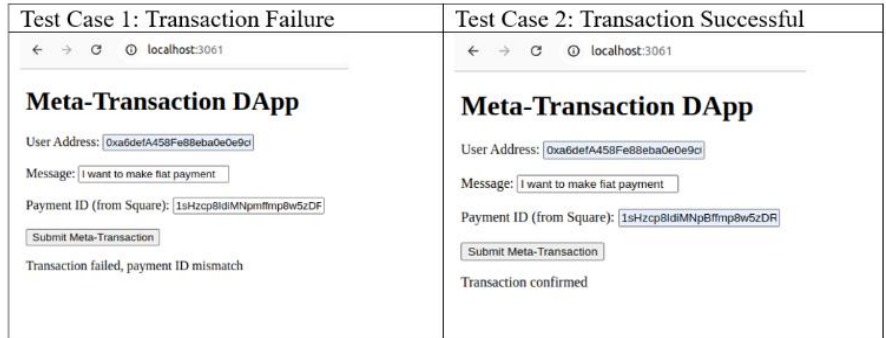

Figure 5 MetaTransaction dApp Output

### 3.4 Performance Evaluation

To evaluate the performance of our proposed gasless subnet-level staking mechanism on the XDC network, we considered both technical and economic performance metrics. These metrics helped us assess the effectiveness, scalability, security, and sustainability of the proposed solution. These are explained below.

A. *Transaction Throughput* - Measures the number of transactions processed per second (TPS) by the gasless subnet. High throughput ensures the network can handle a large volume of transactions, making it suitable for high-frequency applications (e.g., IoT, gaming). The metric used is Transactions per second (TPS).
B. *Latency* - The time it takes for a transaction to be confirmed and finalized on the network. Low latency improves user experience by ensuring fast transaction finalization. The metric used is the average transaction confirmation time (in seconds).
C. *Security and Decentralization* - Measures the network's ability to remain secure and decentralized while operating gasless subnets. It ensures that the network is resistant to attacks and maintains a trustless environment without reliance on a centralized authority. The metric used is number of active validators, distribution of staked XDC across validators (Gini coefficient for decentralization), and rate of successful attack resistance (e.g., 51% attack resilience).
D. *Cost Efficiency (Gasless Model)* - The reduction in transaction costs for users compared to typical gas-based models. One of the main goals of gasless subnets is to reduce the financial burden on users. This metric evaluates the savings achieved by adopting the gasless model. The metric used is Average cost per transaction for users (ideally $0 or near zero) compared to traditional gas-based models.
E. *Validator Incentivization and Sustainability* - Measures the effectiveness of the non-inflationary incentive model for validators. Validators need sufficient rewards to secure the network without relying on inflation. This metric examines how well the redistribution of transaction fees or other non-inflationary rewards works in practice. The metric used is Validator reward per transaction (compared to inflation-based reward systems) and overall validator satisfaction (survey or participation rates).
F. *Scalability* - The ability of the gasless subnet model to scale horizontally (e.g., more subnets) without degrading performance. A scalable solution is critical for supporting the growth of dApps and user base on the XDC Network. The metric used is Number of subnets supported, TPS across subnets, and the impact of subnet proliferation on mainnet performance.

Tables IA through IF show the performance analysis of the proposed solution vs traditional inflation-based model at hypothetical level for the various performance metrics discussed above.

**Table I** Performance Analysis of Gasless vs Traditional Model (Hypothetical)

A. Transaction Throughput

| Model | Congestion | | Impacting Factor |
| --- | --- | --- | --- |
| | Low | High | |
| Traditional | ~1000 | 600-800 | Gas fees prioritization |
| Gasless Subnet | ~800-1000 | 400-600 | Alternative incentivization and prioritization mechanisms |

Traditional model benefits from direct transaction fees and inflationary rewards, which help maintain high throughput even during congestion. In gasless subnet, the throughput is highly dependent on the effectiveness of validator incentives and prioritization methods in the absence of transaction fees. Optimizing these factors is crucial to maintaining high throughput.

B. Latency

| Model | Congestion | | Impacting Factor |
| --- | --- | --- | --- |
| | Low | High | |
| Traditional | ~1-2 s | 5-10 s | Gas fees prioritization, validator participation, and network congestion |
| Gasless Subnet | ~1-3 s | 3-6 s | Validator incentives, transaction volume, subnet performance, cross-subnet/mainnet synchronization |

In low congestion scenarios, the latency of both gasless subnets and traditional inflation-based models should be quite similar, potentially 1-3 seconds depending on the consensus algorithm. In high congestion scenarios, traditional models may experience higher latency due to fee prioritization (e.g., 5-10 seconds), while gasless subnets could see delays around 3-6 seconds due to validator participation issues or lack of transaction prioritization mechanisms. Cross-subnet communication may add slight overhead in gasless models, especially if the subnet relies heavily on interactions with the mainnet.

C. Security and Decentralization

| Model | Impacting Factor | |
|---|---|---|
| | Security (Attack Resistance) | Decentralization |
| Traditional | High, supported by gas fees and inflation rewards | Potential centralization over time as large number of holders accumulates more tokens |
| Gasless Subnet | High, depends on efficient non-inflationary rewards | Higher potential, assuming well-distributed rewards |
| Measurement Method | Total network stake | Validator distribution |

Security is strong in both the models, but relies on the effectiveness of incentivization mechanisms in the gasless model. Decentralization is potentially better in gasless subnets if rewards are distributed evenly, whereas inflation-based models risk centralization over time.

D. Cost Efficiency

| Model | Impacting Factor | | | |
|---|---|---|---|---|
| | User Costs | Validator Revenue | Network Cost | Long-Term Efficiency |
| Traditional | Gas fees (volatile); indirect inflation costs | Gas fees + inflation, consistent returns | Inflation increases supply, token dilution risk | Lower (due to inflation) |
| Gasless Subnet | No gas fees, possible alternative fees (subscription) | Alternative rewards, potentially viable but sustainable | No inflation, stable token value | Higher (no inflation) |

Gasless subnets offer significantly higher cost efficiency for users due to the elimination of gas fees and lower cost volatility. For validators, cost efficiency depends on the alternative reward mechanisms. If well-designed, gasless subnets can be cost-efficient without relying on inflation, but they may introduce more revenue variability compared to traditional models. At the network level, the gasless model is more sustainable in the long run, as it avoids the inflationary pressures that devalue tokens over time, providing a stable and efficient economic model. models risk centralization over time. models risk centralization over time.

E. Validator Incentivization

| Model | Impacting Factor | | | |
|---|---|---|---|---|
| | Rewards Type | Stability of Rewards | Reward Source | Expected Reward/Transaction |
| Traditional | Inflation + Tx fees | High | New tokens minted + user gas fees | Stable (depends on inflation rate and total supply) |
| Gasless Subnet | Redistributed from activity or mainnet staking pool | Moderate | Redistributed fees or other non-inflationary sources | Variable (depends on transaction volume and staking rewards) |

In traditional model, validators are likely to be satisfied in the short term due to predictable, inflation-driven rewards, but there could be dissatisfaction if token inflation leads to devaluation over time. In gasless model, validators may experience greater volatility in rewards, leading to higher satisfaction during periods of high activity and potentially lower satisfaction during low-activity periods. Long-term satisfaction may be higher due to the absence of inflation.

F. Scalability

| Model | Impacting Factor | |
|---|---|---|
| | Number of Subnets Supported | TPS Across Subnets |
| Traditional | Moderate, limited by gas fee competition and inflation effects | ~1000-5000 |
| Gasless Subnet | High, as no gas fees; allow easier scaling of subnets | ~1000-7000 |

Traditional model is moderately scalable in terms of subnets and TPS, but faces limitations due to gas fees and inflationary competition among subnets. Gasless model offers better scalability, supporting more subnets and higher overall TPS, as validators are not influenced by fee prioritization, and the network can operate more smoothly across subnets.

Table II shows the real-time performance analysis of the proposed solution in XDC Apothem Testnet.

**Table II** Real-Time Performance Analysis of the proposed solution in XDC Apothem

| Average Block confirmation time (sec) | Average Cost efficiency | Average Gas price | No. of blocks confirmed (after 20 hrs. of deployment) | Average TPS |
|---|---|---|---|---|
| 2.49 | 0 XDC $ 0.00002433 | 0 XDC (20 Gwei) | 25873 | 8.62 |

The performance is based on a large number of real-time deployments of the proposed dApp (dataset) in the XDC Apothem. Interpretation of the performance is as follows.

- *Confirmation Time:* The transactions are confirmed within 2 to 3 seconds consistently with an average of 2.49 s, indicating that the XDC Apothem network is operating efficiently in terms of transaction finality. This quick confirmation time suggests low network congestion and fast processing of transactions.
- *Cost Efficiency:* Each transaction costs approximately 0 XDC or $0.00002433 (at a gas price of 20 Gwei), which is highly cost-efficient. This makes the network particularly favorable for executing multiple transactions without significant fees.
- *Gas Price:* The gas price is stable at 20 Gwei throughout the dataset, maintaining predictable transaction costs. This stability enhances user confidence in transaction cost predictability, contributing to network efficiency and performance.
- *Block Confirmation:* Block confirmations are steadily increasing, ranging from 25,494 to 25,873. This suggests that while the transaction is confirmed within 2-3 seconds, the number of block confirmations provides additional security over time. More block confirmations ensure the transaction's irreversibility, making the network more secure against potential chain reorganizations.

Analysis of Stress Test of the dApp using 300 virtual users shows the following positive results.

- *Stable Response Times for Small Loads*: For successful requests, the dApp maintains a relatively stable response time, even though it is on the higher end. This suggests that the application can handle low-volume requests reasonably well.
- *Server Stability under Initial Load:* The system did manage to complete 212 out of 300 requests, showing that it can function under a moderate load. This indicates a baseline level of stability and performance that can be optimized further.

The results may be further improved by optimization techniques in future for performance enhancement under stress.

## 4 Use Cases and Applications

### 4.1 Use Cases

The gasless subnet of the XDC Network offers several promising use cases as mentioned below.

- *Microtransactions:* Gasless transactions can facilitate microtransactions by eliminating the need for users to hold and spend tokens for transaction fees. This is particularly useful for small, frequent payments such as those in retail or online services.
- *Credit Networks:* In credit networks, gasless transactions can streamline processes by reducing the overhead costs associated with transaction fees. This can make credit services more accessible and affordable for users.
- *Peer-to-Peer Payments:* Gasless subnets can enhance peer-to-peer payment systems by making transactions more seamless and user-friendly. Users can send and receive payments without worrying about transaction fees, which can encourage more frequent use.
- *Subscription Services:* For subscription-based services, gasless transactions can simplify the payment process. Users can subscribe to services without needing to manage transaction fees, making the experience smoother and more attractive.
- *Loyalty Programs:* Businesses can implement loyalty programs using gasless transactions, allowing customers to earn and redeem points without incurring additional costs. This can increase customer engagement and satisfaction.

These use cases highlight how gasless subnets can improve efficiency, reduce costs, and enhance user experience in domestic payment and credit networks. There are several real-world examples of successful gasless systems. The 0x Protocol offers a Gasless API that allows users to perform ERC-20 token swaps without needing to pay gas fees directly. Instead, a third party covers the gas costs, providing a seamless experience for users [6]. Gelato Network enables developers to create gasless dApps. For example, they have implemented gasless voting for DAOs, allowing members to vote on proposals without paying gas fees. This increases participation by removing the friction associated with gas costs [7]. Biconomy provides a gasless transaction infrastructure that allows dApps to offer a smooth user experience by abstracting away gas fees. This is particularly useful for onboarding new users who might be unfamiliar with managing gas fees. These examples demonstrate how gasless systems can enhance user experience and increase participation by removing the barrier of transaction fees.

### 4.2 Applications

Blockchain and federated learning (FL) can work together to create a secure, decentralized, and privacy-preserving environment for machine learning. FL is a machine learning approach where multiple devices or servers collaborate to train a model without sharing their raw data. Instead, each participant trains the model locally and only shares the model updates (gradients) with a central server, which aggregates them to improve the global model. This approach enhances data privacy and security since the raw data never leaves the local devices. Blockchain is a decentralized ledger technology that ensures data integrity, transparency, and security through cryptographic techniques. It records transactions in a tamper-proof manner, making it ideal for applications requiring trust and accountability. There are several benefits of integrating blockchain to FL.

- *Data Integrity and Trust:* Blockchain ensures that the model updates shared by participants in FL are tamper-proof and authentic. Each update can be recorded on the blockchain, providing a transparent and immutable audit trail.
- *Decentralization:* Both technologies are inherently decentralized. FL distributes the training process across multiple devices, while blockchain eliminates the need for a central authority. This synergy enhances the robustness and resilience of the system.
- *Incentive Mechanisms:* Blockchain can implement smart contracts to incentivize participants in FL. For example, participants can be rewarded with tokens for contributing high-quality model updates, encouraging active and honest participation.
- *Privacy Preservation:* FL already enhances privacy by keeping raw data local. Blockchain adds an extra layer of security by ensuring that the model updates are securely and transparently recorded, preventing malicious actors from tampering with the data.
- *Regulatory Compliance:* Blockchain's transparency and immutability can help in meeting regulatory requirements, such as data provenance and auditability, which are crucial in sectors like healthcare and finance.

Figure 6 shows an example workflow of this integration. There are five blocks each having its own roles and responsibilities.

- *Local Training:* Each participant trains the model on their local data.
- *Model Update:* Participants send their model updates to a blockchain network.
- *Verification:* The blockchain network verifies the authenticity and integrity of the updates.
- *Aggregation:* A smart contract aggregates the verified updates to improve the global model.
- *Incentives:* Participants are rewarded for their contributions through blockchain-based tokens.

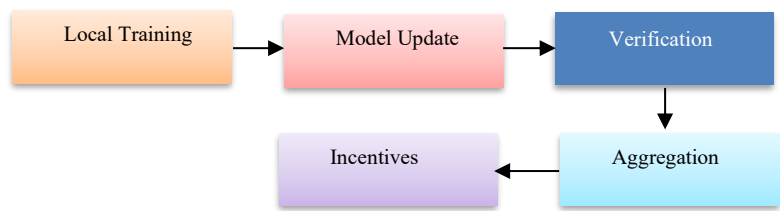

Figure 6     Example Workflow of Blockchain and FL Integration

This integration creates a robust framework for collaborative machine learning while ensuring data privacy, security, and trust [8]. Integrating a gasless subnet of the XDC Network into FL models can offer several innovative use cases.

- *Cost-Efficient Model Training:* By eliminating transaction fees, a gasless subnet can make FL more cost-effective, especially when frequent model updates and communications are required [9].
- *Enhanced Privacy and Security:* FL inherently focuses on data privacy by keeping data localized. Integrating with a gasless subnet can further enhance security by ensuring that transactions (such as model updates) are secure and tamper-proof [10].
- *Scalability:* Gasless subnets can handle a high volume of transactions without the bottleneck of gas fees, making it easier to scale FL models across numerous devices and clients [11].
- *Regulatory Compliance:* FL models often need to comply with data protection regulations. A gasless subnet can help ensure that all transactions are transparent and auditable, aiding in regulatory compliance [12].
- *Incentivization:* Participants in FL can be incentivized through token rewards without worrying about gas fees, encouraging more participation and data sharing [13].

FL models integrated with blockchain technology are being explored and implemented in various countries, each with unique approaches tailored to their specific needs and regulatory environments. Researchers in China have been actively working on blockchain-empowered federated learning (BC-FL) models to enhance data privacy and security. These models are particularly useful in sectors like healthcare, where patient data privacy is paramount. The integration of blockchain helps in creating a secure and transparent data-sharing environment [14]. In the U.S., blockchain-based federated learning models are being explored for applications in finance and healthcare. These models leverage blockchain to ensure data integrity and provide a decentralized approach to data sharing, which is crucial for complying with stringent data privacy regulations [15]. The EU has stringent data protection laws, such as the General Data Protection Regulation (GDPR). Federated learning models in the EU often incorporate blockchain to enhance compliance with these regulations. Blockchain provides an immutable ledger that ensures transparency and accountability in data processing [16]. Singapore is focusing on blockchain-based federated learning models for smart city applications. These models help in managing and analyzing data from various IoT devices while ensuring data privacy and security. Blockchain aids in creating a trustless environment where data can be shared securely among different entities.

# 5 The Challenges

The challenges of implementing a gasless subnet architecture on the XDC blockchain, or any similar network, revolve around both technical and economic factors Let us delve deep into the details and the corresponding probable solutions.

- *Economic Incentives:* Validators typically earn transaction fees (gas fees) as a reward for maintaining and securing the network. In a gasless subnet, the lack of transaction fees may reduce the incentive for validators to participate unless alternate reward structures (e.g., staking rewards, or fiat-based payments) are devised. If alternative rewards are based on inflationary token issuance (e.g., minting more tokens), this can devalue the network's native token unless well-managed. An alternative reward structure can be established using a combination of staking rewards and fiat-based payments. These rewards can be sourced from transaction fees collected in fiat or from a decentralized treasury funded by network activities. To avoid inflationary token issuance, which could devalue the native token, the reward system can be designed to redistribute existing tokens or utilize non-inflationary incentives, such as staking bonuses or performance-based rewards. This ensures validators remain incentivized while preserving the value of the native token.

- *Security Considerations:* In traditional blockchain models, gas fees act as a deterrent to spam transactions and denial-of-service attacks. Without fees, the network is more vulnerable to attack unless other mechanisms (like rate-limiting or proof-of-work/proof-of-stake validation) are implemented. Subnets operating without gas need alternative ways to prevent Sybil attacks, where an attacker could create numerous nodes or addresses to flood the network with fraudulent activity. Security mechanisms such as rate-limiting, proof-of-stake (PoS), or proof-of-authority (PoA) validation can be implemented. Rate-limiting would cap the number of transactions per user or address, while PoS or PoA ensures that only validators with sufficient stakes or reputation can validate transactions, making it costly for attackers to flood the network. Additionally, identity-based mechanisms like decentralized identity (DID) or reputation systems could be used to restrict malicious actors from creating numerous fraudulent nodes, thereby ensuring robust network security without gas fees.

- *Fiat Fee Economic Model:* Designing an economic model where transaction fees are paid in fiat currency rather than a native token requires an efficient fiat on-ramp and off-ramp mechanism. This could introduce complexities like regulatory compliance, banking relationships, and fluctuating exchange rates. As subnets might operate in different jurisdictions, managing cross-border fiat payments for fees could become complex, particularly in terms of currency conversion, compliance, and transaction speed. An efficient fiat on-ramp and off-ramp mechanism is essential, integrating licensed payment gateways and banking partnerships to facilitate smooth currency transactions. Stablecoins or fiat-pegged assets can be used to mitigate exchange rate fluctuations, ensuring stable value transfers across subnets. Additionally, decentralized finance (DeFi) protocols can provide liquidity for cross-border currency conversions, and compliance with regional regulations can be handled through automated AML/KYC checks embedded in the payment process. This ensures secure, compliant, and efficient fiat transactions across jurisdictions, while maintaining transaction speed and minimizing complexities.

- *Network Complexity:* Gasless subnets will need to interoperate with other XDC subnets and the main XDC chain. Without uniform gas fees, transactions between gasless and gas-based subnets could become complicated in terms of state verification and validation. Gas fees typically play a role in the consensus mechanism. In a gasless network, the consensus model may need to be modified to ensure robust security, decentralization, and validator engagement. A modified consensus model that decouples gas fees from validation while maintaining security and decentralization is needed. This could involve using a cross-chain bridge that standardizes transaction validation and state verification between the two types of subnets, leveraging transaction fees or incentives from the gasless network's fiat-based payments. Validators on the gasless subnet can be rewarded through mechanisms like staking or fee-sharing from the network's treasury, ensuring they remain engaged without relying on traditional gas fees. This ensures secure, scalable, and efficient cross-subnet transactions.

- *Long-Term Sustainability:* Running a gasless subnet still involves infrastructure costs (such as storage, bandwidth, and computation). If users are not directly paying for these costs through gas, the question arises: who bears the costs in the long term? To sustain the gasless subnet, some entity (e.g., the project creator or a decentralized treasury) must subsidize operational costs, raising concerns about sustainability as the network scales. A decentralized treasury model can be implemented, where fees from fiat transactions or staking rewards are allocated to subsidize operational expenses such as storage, bandwidth, and computation. Additionally, innovative mechanisms like fee-sharing, sponsorships, or value-added services (e.g., premium features for businesses or high-frequency users) can help generate ongoing revenue. This model can be supported by governance, where participants vote on how to allocate funds, ensuring the network remains scalable and sustainable without relying solely on project creators.

- *User Adoption and Education:* While a gasless system simplifies user experience (since users don't need to hold tokens to interact), educating users on this novel architecture, particularly around fiat-based payments and transaction models, may require effort. Developers may need to adjust dApps to fit the new gasless model, as most are designed with gas fee mechanisms in mind. A multi-pronged strategy is needed. This involves creating clear, accessible educational resources to guide users through the new transaction model, focusing on the benefits of simplified payments without the need for tokens. Additionally, developers can provide tools and SDKs that abstract the gasless architecture, allowing seamless integration of existing dApps into the new model. Incentivizing developers to adapt their dApps by offering grants or rewards can accelerate the transition, while user-friendly interfaces and tutorials ensure smooth adoption of the fiat-based transaction experience.

- *Regulatory Compliance:* Subnets operating with fiat fees must comply with relevant financial regulations, including Anti-Money Laundering (AML) and Know Your Customer (KYC) requirements, potentially complicating the decentralized ethos of blockchain. Governance mechanisms will need to address issues related to fee redistribution, economic incentives, and validator rewards. Decision-making on such matters in a decentralized environment can be slow and contentious. A hybrid governance model combining on-chain mechanisms like DAOs with off-chain regulatory compliance can be implemented. This approach would use middleware solutions for AML/KYC verification, leveraging cryptographic techniques like Zero-Knowledge Proofs (ZKPs) to preserve user privacy. Additionally, smart contract-based fee redistribution and dynamic incentives for validators would streamline economic rewards and decision-making. Delegated Proof of Stake (DPoS) can enhance governance efficiency by empowering trusted validators, while compliance nodes or Compliance-as-a-Service (CaaS) handle regulatory tasks, ensuring rapid, compliant, and decentralized operation.

All of the above-mentioned challenges must be carefully considered to ensure a smooth and secure implementation of a gasless subnet within the XDC network. Table III depicts the challenge-solution (C-S) analysis for the *chosen* impact.

## 6 The Roadmap

The future roadmap for the gasless subnet architecture on the XDC network may encompass several strategic phases to ensure the gasless subnet's growth, security, and sustainability while adapting to evolving market and regulatory landscapes.

The network may enhance transaction throughput and network performance through ongoing optimization and the integration of Layer-2 solutions for scaling, develop robust interoperability mechanisms to facilitate seamless transactions between the gasless subnet and other blockchain ecosystems i.e., undergoing cross-chain integration, expand developer outreach with grants, hackathons, and comprehensive tools to encourage dApp development on the gasless subnet, focus on integrating enterprise solutions, demonstrating real-world use cases that benefit from gasless transactions and fiat payments, forge partnerships with additional payment providers to

**Tabel III**     Challenge-Solution Analysis with Impact

| Challenge (C) | Solution (S) | C-S Analysis | Impact *(Chosen)* |
|---|---|---|---|
| Economic Incentives | Alternative reward structure | This avoids inflationary token issuance and maintains the value of the native token | High |
| Security Considerations | Rate-limiting | Limit attacker access, and restrict fraudulent nodes | Medium |
| Fiat Fee Economic Model | Licensed payment gateways | ensure secure, compliant, and efficient transactions across jurisdictions, maintaining transaction speed and minimizing complexities | High |
| Network Complexity | Modified consensus model | Decouple gas fees from validation while maintaining security and decentralization | High |
| Long-Term Sustainability | Decentralized treasury model | Ensures network scalability and sustainability without relying solely on project creators | Medium |
| User Adoption and Education | Comprehensive strategy | Facilitates the fiat-based transaction experience | Low |
| Regulatory Compliance | Hybrid governance model | Enhances governance efficiency and ensures decentralized operation | High |

offer more fiat on-ramp and off-ramp options across diverse jurisdictions, implement AI-based systems for enhanced spam and Sybil attack prevention and continuously audit network security.

At the same time the network may keep up with global regulatory changes, ensuring compliance in multiple regions with updated AML/KYC protocols, broaden DAO functionalities for decentralized decision-making on network governance, treasury management, and validator rewards, explore new revenue models, such as premium services and staking derivatives, to fund the network sustainably by adopting federated learning AI models.

There comes an urgent need to adjust tokenomics to balance rewards and incentives without relying on inflationary practices and target use cases in emerging markets where gasless transactions and fiat payments provide significant benefits such as DeFi and NFT for fostering growth, attracting creators and investors with gasless transaction advantages.

Another important aspect is to explore FL based innovative ideas leveraging them to XDC network whereby content creators that get advertisement revenues can use FL model to further improve their reach. This would enable continuous refinement of the network's performance based on real-world data and usage patterns, ensuring the network meets user needs and expectations. Furthermore, providing resources and training for users and developers to facilitate the adoption and effective use of the gasless subnet must be considered for developing long-term strategies for environmental and operational sustainability within the network infrastructure.

## 7 The Team

The team comprises a variety of skilled professionals who have collaborated to research, design, development, testing, and maintenance of the gasless subnet staking proposed system in the XDC network.

*Mohuya Chakraborty* is the blockchain engineer who conceived the idea, designed the overall structure of the blockchain system. *Sudip Kumar Palit* is the blockchain and smart contract developer who focused on the core development and maintenance of the blockchain and smart contract, and design of the UI/UX for the blockchain dApp in XDC Apothem. *Pramod Viswanath, Gerui Wang and Fisher Yu* are the blockchain researchers who tested the blockchain system in XDC Mainnet to ensure they are free of bugs and functions as intended. *Atul Khekade* is the project director who supervised the project to ensure it stayed on track and met the deadline of the project.

## 8 Conclusion

The dApp on the XDC network that serves as a gasless subnet staking mechanism has been proposed in this paper. To spin off a subnet that operates like a non-crypto network and accepts currency fees on the XDC network, the strategy is to build a gasless subnet inside an XDC mainnet. This would increase scalability and make staking more economical and efficient by enabling users to stake their tokens without paying gas fees. Promising results were obtained when the dApp was tested on Apothem XDC in terms of throughput (8.62 TPS), block confirmation time (2.49s), cost ($0.00002433), and gas price (20 Gwei). As a result, it performs admirably in terms of swift transaction

finality, extremely cheap fees, and stable gas prices. In addition, block confirmations are continuously raised to improve security (tested after 20 hours of deployment). This degree of security and efficiency is perfect for those who want blockchain transactions that are quick and affordable. Based on its current performance, it also shows good potential for scaling. This keeps speed, cost, and security intact while putting the network in a good position for future expansion. The use cases and applications of the present work along with challenges and probable solutions are expressed in a lucid manner. The future prospects of the proposed solution on the XDC network are explained for long-term sustainability of the model